\documentclass[12pt]{article}
\usepackage{blindtext}
\usepackage{titlesec}
\usepackage{graphicx}
\usepackage{tcolorbox}
\usepackage{mdframed}
\usepackage{pdfpages}
\usepackage{epstopdf}
\AppendGraphicsExtensions{.pdf}
\usepackage{subfig}
\usepackage{amssymb, amsmath, amsthm}
\usepackage{graphicx}
\usepackage{cite}

\newcommand{\const}{\,{\rm const}\,}

\usepackage{hyperref}

\def\be{\begin{equation}}
	\def\ee{\end{equation}}
\def\bea{\begin{eqnarray}}
	\def\eea{\end{eqnarray}}

\theoremstyle{definition}

\usepackage[left=2cm,right=2cm,top=2cm,bottom=2cm]{geometry}

\title{}
\author{}

\begin{document}
\title{Distorted black ring with rotation on the two-sphere}
\author{Matin Tavayef$^a$\footnote{mtavayef@mun.ca}, Shohreh Abdolrahimi $^b$\footnote{sabdolrahimi@cpp.edu} , Ivan Booth$^c$\footnote{ibooth@mun.ca}, and Hari Kunduri$^d$\footnote{kundurih@mcmaster.ca}  \\ \\
\small \sl $^a$Faculty of Science, Theoretical Physics, Memorial University of Newfoundland \\ \small \sl  St. John's, NL  A1C 5S7, Canada \\
\small \sl $^b$Department of Physics and Astronomy, California State Polytechnic University, \\ \small \sl 
Pomona, CA 91768, USA  \\
\small \sl $^c$Department of Mathematics and Statistics, Memorial University of Newfoundland \\ \small \sl 
St. John's, NL,  A1C 5S7, Canada  \\
\small \sl $^d$Department of Mathematics and Statistics \& Department of Physics and Astronomy \\ \small \sl  McMaster University,  Hamilton, ON, L8S 4M1, Canada } 

\maketitle

\begin{abstract}
We construct a stationary black ring spacetime with rotation only along the $S^2$ direction that is free of conical singularities. We use a solution-generating technique to add external gravitational fields and show these can be used to remove conical singularities from an asymptotically flat black ring with a rotating $S^2$. 
\end{abstract}

\section{Introduction}
In four-dimensional general relativity, black hole uniqueness theorems impose strong restrictions on the space of stationary, asymptotically flat (analytic) solutions of the Einstein equations. Specifically, the Kerr family of solutions exhausts the set of possible solutions, and hence vacuum black holes are completely characterized by their mass and angular momentum, ahd have horizons of spherical topology \(S^2\)~\cite{israel1967,carter1971,robinson1975, Chrusciel:2012jk}.  However, this uniqueness breaks down in higher dimensions, revealing a much richer space of gravitational solutions (see, e.g. \cite{Emparan:2008eg} for a review). A particular example of this is the existence of five-dimensional \emph{black rings}, which are black holes with horizon topology of \(S^1 \times S^2\)~\cite{emparanreall2002}.

The original two-parameter family of vacuum black ring solutions, constructed by Emparan and Reall~\cite{emparanreall2002}, describes an asymptotically flat vacuum black hole that is smooth on and outside the event horizon. The horizon is prevented from collapse by carrying angular momentum in the $S^1$ direction. Subsequently, this solution was shown to be a subset of a larger family of black ring vacuum solutions which carry independent angular momenta along both the $S^1$ and $S^2$ directions~\cite{pomeransky2006}. The original black ring solution was also generalized to include dipole charges~\cite{emparan2004}. Supersymmetric (those with Killing spinors) families of supergravity solutions have also been constructed~\cite{elvang2004}. However, the set of solutions with non-vanishing angular momenta only along the $S^2$ direction, originally constructed by Figueras \cite{Figueras}, necessarily have conical singularities along at least one of the poles of the $S^2$, as is the case for the static black ring. Physically, this suggests that, at least in the vacuum, rotation on the $S^2$ direction of the ring is insufficient to prevent collapse. 

The goal of this work is to construct a regular, local $S^2-$rotating black ring vacuum solution. The local rotating black ring we shall present in this paper can be interpreted as a \emph{distorted black hole}, solutions that describe black holes deformed under the influence of external gravitational fields. The foundational work of Geroch and Hartle~\cite{gerochhartle1982} proposed that certain asymptotically non-flat vacuum solutions known as distorted black holes are local solutions to the Einstein equations that could be interpreted physically as black holes subject to the gravitational infleunece of external matter distributions. 

In four dimensions, early examples included the Erez–Rosen solution~\cite{erezrosen1959}, which added a quadrupole distortion to the Schwarzschild metric, and subsequent extensions to the charged and rotating cases~\cite{breton1997, fairhurst2001, yazadjiev2001}. These studies demonstrated that distorted black holes retain many essential properties of their highly constrained isolated counterparts, while some features of these black holes are different. 

Theoretical investigations of distorted black holes and black objects are important for understanding the broader landscape of solutions in general relativity. Unlike idealized, isolated spherically symmetric black holes, distorted/deformed configurations represent more general static or stationary spacetimes that incorporate the influence of external gravitational fields. They provide a valuable framework for probing how key features of black holes—such as horizon structure, thermodynamic relations, and geometric inequalities—are modified under deformations. Our goal is to gain more insight into the question of which properties of black holes are truly generic. Initial analyses of distorted black holes focused on static, axisymmetric distortions of Schwarzschild black holes, demonstrating that external multipole fields can deform the horizon geometry and affect observable properties of the black hole spacetime \cite{israelkhan1964, doroshkevich1965, erezrosen1959}. Later developments extended these methods to include rotating black holes, particularly through the application of solitonic solution-generating techniques. These approaches enabled the construction of distorted stationary black hole solutions that retain axisymmetry, but exhibit modified mass, angular momentum, and ergoregion structure due to the external field \cite{tomimatsu1984, breton1997}.

Analysis of these solutions has revealed many speical  properties of distorted black holes, as well as illustrating that many of the structural and thermodynamic features known from isolated black holes carry over to the distorted case. Notably, the Smarr formula remains valid when defined locally on the horizon \cite{breton1998, papadopoulos1984, hennig2008, hennig2010, ansorg2011}. 
The interior structure of distorted black holes has also been studied, \cite{frolovshoom2006, abdolrahimi2009}. Perhaps most strikingly, it has been shown that distortion allows for a violation of extremality bounds characteristic of isolated black holes. In particular, the angular momentum-to-mass ratio can exceed the usual bound \(|J|/M^2 \leq 1\) without destroying the horizon, and in some cases can grow arbitrarily large \cite{ansorg2005, AbdolrahimiKerr}. 

The distorted black hole paradigm has since been extended to five-dimensional black objects \cite{Abdolrahimi2010} using the generalized Weyl formalism \cite{EmparanReall}. In this approach, the general solution of the $D$-dimensional vacuum Einstein equations that admits $D-2$ mutually orthogonal commuting non-null Killing vector fields is given either in terms of $D-3$ independent axisymmetric solutions of Laplace's equation in three-dimensional flat space or by $D-4$ independent solutions of Laplace's equation in two-dimensional flat space. This underlying linear structure enables, by superposition, the addition of external multipole moments that distort the background solution without violating the Einstein equations (the expicit solution is known up to quadrature). Using the same method, a distorted 5-dimensional charged Shwarzschild-Tangherlini black hole \cite{Abdolrahimi2014}, a distorted black ring \cite{Abdolrahimi2020} and a distorted static black hole with a bubble (a non-trivial $S^2$ in the black hole exterior) has been constructed \cite{Tavayef2022}. It was shown that the strength of the external sources can be adjusted to eliminate the conical singularities that necessarily arise in the asymptotically flat static black ring solution or the asymptotially flat static black hole with a bubble. This demonstrates the critical role of distortion in achieving equilibrium configurations that are otherwise unavailable in solutions taht describe isolated vacuum systems. 

The generalized Weyl formalism cannot be directly used in the case of stationary, non-static spacetimes because the timelike Killing field is necessarily not hypersurface-orthogonal. In this case, we may use powerful \emph{solitonic generating techniques}, including the inverse scattering method and B\"acklund transformations. The B\"acklund transformation, originally developed by Neugebauer in the context of four-dimensional stationary axisymmetric spacetimes~\cite{neugebauer1979}, was adapted to five dimensions for generating rotating solutions from static seeds~\cite{iguchi2006}. For example, the \( S^2 \)-rotating black ring \cite{Figueras} was systematically constructed by Iguchi and Mishima using a 2-fold B\"acklund transformation acting on a static Weyl-class seed \cite{iguchi2006}. Recent developments illustrate that solution-generating techniques may have deeper physical meaning and are not merely just a trick \cite{Joon-HwiKim}. For example, it was shown that the Newman-Janis algorithm precisely originates because the Kerr metric represents a pair of self-dual and anti-self-dual gravitational dyons (Taub-NUT instantons). 

Combining the generalized Weyl formalism and the above-mentioned solitonic generating techniques with the distorted black hole formalism, a distorted Myers-Perry black hole with a single angular momentum was constructed starting from a distorted Minkowski background seed solution \cite{abdolrahimi2014B}. 

In this work, we also bring these threads together to construct a new class of solutions that represent a distorted rotating black ring. We start with an appropriately `distorted Minkowski seed solution and then apply a similar set of solution-generating steps following \cite{iguchi2006} and \cite{iguchi2006} to produce the desired local solution. We then analyze various special cases to show that it is possible to obtain smooth black ring metrics with rotation only along the $S^2-$direction without any conical singularities. 

The work is organized as follows. In Section 2 we review the solitonic solution generating technique for constructing stationary five-dimeisional  $\mathbb{R} \times U(1) \times U(1)$-invariant solutions. In Section 3 we apply this technique to construct the local solution and show how the parameters of the solution can be chosen so that the resulting spacetime describes a black ring with rotation  along the $S^2-$direction that is free of conical singularities. In Section 4, we consider some properties of the solution and conclude with a brief discussion. 

\section{Solitonic solution-generating technique} It is well known that the four-dimensional vacuum Einstein equations reduced on stationary and axisymmetric solutions can be recast as a pair of coupled nonlinear equations. In turn, these equations can be reformulated in terms of a single equation for a complex Ernset potential. Moreover, this equation is integrable and thus can be solved explicitly using solution-generating methods.   The five-dimensional vacuum equations reduced on stationary and bi-axisymmetric solutions (admitting a $\mathbb{R} \times U(1) \times U(1)$ group of isometries) can also be formulated in this way when one of the $U(1)$ Killing fields is hypersurface-orthogonal. Here we briefly review this formulation, which is given in greater detail in \cite[Section II]{iguchi2006}.

We begin with the metric of a stationary, bi-axisymmetric metric where one of the spatial Killing fields is hypersurface-orthogonal.  This may always be written in the Weyl-Papapetrou coordinates $(x_0, \psi,\phi, \rho,z)$:\cite{Halmark}
\begin{equation}\label{WP}
	ds^{2}=-e^{2U_{0}}(dx_{0}-\omega d\phi)^{2}+e^{2U_{1}}\rho^{2}(d\phi)^{2}+e^{2U_{2}}(d\psi)^{2}+e^{2(\gamma+U_{1})}( d\rho^{2}+dz^{2})
\end{equation} where $x_0$ is a timelike coordinate associated with the stationary isometry generated by $\partial_{x_0}$ and $(\psi,\phi)$ are coordinates associated to the $U(1) \times U(1)$ action (for the moment we do not assume any identifications in the $(\psi,\phi)$ plane). The remaining coordinates $(\rho,z)$ may be thought of as global coordinates on the space of orbits $B = M / (\mathbb{R} \times U(1)^2)$. $B$ can be identified with the upper half plane $\{(\rho,z)| \rho \geq 0, z \in \mathbb{R}\}$~\cite{Hollands:2007aj}. The metric is determined by the functions $ U_{0} $, $   U_{1} $, $  U_{2}  $, $ \omega $ and $ \gamma $ which are depend only on $ \rho $ and $ z $. In particular, when $\omega=0$ the vacuum equations impose that $U_i$, $i = 0,1,2$ are axisymmetric, harmonic functions on $\mathbb{R}^3$ away from the $z-$axis. The boundary of $B$ corresponds to $\rho =0$. It can be observed that $\rho =0$ if and only if the restriction of the metric to the Killing fields is degenerate. This occurs when one or both of the spatial Killing fields degenerate (referred to as axes and corner points, respectively) or a timelike Killing field becomes null. The number of rotation axes and event horizons in the spacetime is therefore determined by the zero set of $\rho$. 

By defining functions $  S $ and $ T $ as $ S:=2 U_{0}+ U_{2}  $ and $ T:= U_{2}  $, the above metric can be written as
\begin{equation}\label{WPST}
	ds^{2}=e^{-T}[-e^{S}(dx_{0}-\omega d\phi)^{2}+e^{T+{2U_{1}}}\rho^{2}(d\phi)^{2}+e^{2(\gamma+U_{1})+T}( d\rho^{2}+dz^{2})]+e^{2T}(d\psi)^{2}
\end{equation}
By using this metric, the Einstein equations simplify to the following equations for the metric functions $(T, U_1, S, \gamma, \omega)$:
\begin{gather}\label{EFE}
	\nabla^{2} T=0, \qquad U_{1}=-\dfrac{S+T}{2} \\
\gamma=\gamma_{S}+\gamma_{T} \\
\partial_{\rho} \gamma_{T}=\frac{3\rho}{4}[(\partial_{\rho}T)^2-(\partial_{z}T)^2], \qquad
\partial_{z} \gamma_{T}=\frac{3\rho}{2}[(\partial_{\rho}T)(\partial_{z}T)] \\
\partial_{\rho} \gamma_{S}=\frac{\rho}{2(\varepsilon_{S}+\bar{\varepsilon}_{S})}\left(\partial_{\rho}\varepsilon_{S}\partial_{\rho}\bar{\varepsilon}_{S}-\partial_{z}\varepsilon_{S}\partial_{z}\bar{\varepsilon}_{S}\right) \qquad 
		\partial_{z} \gamma_{S}=\frac{\rho}{2(\varepsilon_{S}+\bar{\varepsilon}_{S})}\left(\partial_{\rho}\varepsilon_{S}\partial_{z}\bar{\varepsilon}_{S}+\partial_{z}\varepsilon_{S}\partial_{\rho}\bar{\varepsilon}_{S}\right),
\end{gather} and 
\begin{equation}\label{Ernest}
	\nabla^{2} \varepsilon_{S}=\frac{2}{\varepsilon_{S}+\bar{\varepsilon}_{S}}\nabla \varepsilon_{S}.\nabla\varepsilon_{S},
\end{equation}
\begin{equation}\label{phi}
	(\partial_{\rho}\Phi,\partial_{z}\Phi)=\rho^{-1}e^{2S}(-\partial_{z}\omega,\partial_{\rho}\omega),
\end{equation}
where the complex function $ \varepsilon_{S} $ is defined by $ \varepsilon_{S}:=e^{S} + i\Phi$ and $ \Phi $ can be determined up constants of integration by \ref{phi}. The two-dimensional equation (\ref{Ernest}) is exactly the same as the four dimensional Ernest equation \cite{Ernest}, allowing us to refer to $ \varepsilon_{S} $ as the Ernst potential. Note that $\rho^2 = -\det G$ where $G_{ij} = g(\xi_i, \xi_j)$, where is the Killing part of the metric as discussed above, with $\xi_i$ denoting the Killing vector fields associated to the coordinates $\{x_0, \psi,\phi \}$. 

The integrability of the first-order equations for $\gamma_T$ and $\gamma_S$ is guaranteed as a consequence of the remaining Einstein equations. To obtain explicit forms of the metric functions, we will employ the method presented by Castejón-Amenedo and Manko \cite{Manko90}, who investigated deformations of a Kerr black hole under the influence of external gravitational fields.

The solution-generating technique we use to construct our solution is to distort a static seed solution and then apply a B\"acklund transformation to produce a stationary, non-static vacuum solution. 
The form of a general static  ($\omega_0 =0$) and bi-axisymmetric seed metric is given by
\begin{equation}\label{WPstatic}
	ds^{2}=e^{-T_{0}}[-e^{S_{0}}(dx_{0})^{2}+e^{-S_{0}}\rho^{2}(d\phi)^{2}+e^{2\gamma_{0}-S_{0}}( d\rho^{2}+dz^{2})]+e^{2T_{0}}(d\psi)^{2}
\end{equation} The associated Ernst potential is simply $e^{S_0}$. 
Under the B\"acklund transformation, a new Ernst potential can be written in the form
\begin{equation}
\varepsilon_{S}=e^{S_{0}}\dfrac{x(1+ab)+iy(b-a)-(1-ia)(1-ib)}{x(1+ab)+iy(b-a)+(1-ia)(1-ib)}
\end{equation}
where $x$ and $y$ are prolate-spheroidal coordinates defined implicitly by 
\begin{equation} \label{rhoz}
\rho=\sigma \sqrt{x^{2}-1} \sqrt{1-y^{2}}, \qquad z=\sigma xy 
\end{equation} such that $ \sigma >0 $, $ 1 \leq x $ and $ -1 \leq y \leq 1 $. 
The following first-order differential equations are satisfied by the functions $a=a(x,y)$ and $b=b(x,y)$:
\begin{eqnarray}\label{diffa-gl}
	(x-y)\partial_{x}a=a[(xy-1)\partial_{x} S_{0}+(1-y^{2})\partial_{y}S_{0}],\\ \nonumber
	(x-y)\partial_{y}a=a[-(x^{2}-1)\partial_{x} S_{0}+(xy-1)\partial_{y}S_{0}]
\end{eqnarray}
\begin{eqnarray}\label{diffb-gl}
		(x+y)\partial_{x}b=-b[(xy+1)\partial_{x} S_{0}+(1-y^{2})\partial_{y}S_{0}],\\ \nonumber
	(x+y)\partial_{y}b=-b[-(x^{2}-1)\partial_{x} S_{0}+(xy+1)\partial_{y}S_{0}]
\end{eqnarray}
In five dimensions, the metric functions are \cite{Manko90},
\begin{equation}
	e^{S}=e^{S_{0}}\dfrac{A}{B}, \quad \omega=2\sigma e^{-S_{0}}\dfrac{C}{A}+C_{1},\quad e^{\gamma}=C_{2}(x^{2}-1)^{-1}Ae^{2\gamma^{\prime}},
\end{equation}
where $ C_{1} $ and $ C_{2} $ are constants and $ A, B $ and $ C $ are given by
\begin{equation}\label{A}
	A:=(x^{2}-1)(1+ab)^{2}-(1-y^{2})(b-a)^{2},
\end{equation}
\begin{equation}\label{B}
	B:=[(x+1)+(x-1)ab]^{2}+[(1+y)a+(1-y)b]^{2},
\end{equation}
\begin{equation} \label{C}
	C:=(x^{2}-1)(1+ab)[b-a-y(a+b)]+(1-y^{2})(b-a)[1+ab+x(1-ab)],
\end{equation}
Th function $\gamma^{\prime}$ is a $ \gamma $ function corresponding to the static metric 
\begin{align}\label{static2}
	ds^{2}&=e^{-T_{(0)}}[-e^{(2U_{0}^{(BH)}+S_{(0)})}(dx_{0})^{2}+e^{-2U_{0}^{(BH)}-S_{(0)}}\rho^{2}(d\phi)^{2}+e^{2(\gamma^{\prime}-U_{0}^{(BH)})-S_{0}}( d\rho^{2}+dz^{2})]\\
	& +e^{2T_{0}}(d\psi)^{2} \nonumber
\end{align}
where
\begin{equation}\label{UBH}
	 U_{0} ^{(BH)}=\frac{1}{2}\ln \left(\dfrac{x-1}{x+1}\right).
\end{equation}
 The function $T=T_{0}$ and $U_{1} $ is determined by \eqref{EFE}. The function $\gamma^{\prime} $ satisfies the following equations
\begin{equation}\label{diffr-Gprime}
\partial_{\rho}\gamma^{\prime}=\frac{\rho}{4} \left[(\partial_{\rho}S^{\prime})^{2}-(\partial_{z}S^{\prime})^{2}\right]+\frac{3\rho}{4} \left[(\partial_{\rho}T^{\prime})^{2}-(\partial_{z}T^{\prime})^{2}\right]
\end{equation}
\begin{equation}\label{diffz-Gprime}
	\partial_{z}\gamma^{\prime}=\frac{\rho}{2} \left[\partial_{\rho}S^{\prime}\partial_{z}S^{\prime}\right]+\frac{3\rho}{2} \left[\partial_{\rho}T^{\prime}\partial_{z}T^{\prime}\right]
\end{equation}
where $ S^{\prime} $ and $ T^{\prime} $ can be read out from (\ref{static2}) as
\begin{equation}\label{Tp,Sp}
	S^{\prime}=2U_{0}^{(BH)}+S_{0},\quad T^{\prime}=T_{0}.
\end{equation} First order equations of the type \eqref{diffr-Gprime} and \eqref{diffz-Gprime} can be integrated explicitly (see e.g. \cite[Eq. (22)]{iguchi2006}). We will carry this out in our particular case below to explicitly construct our solutions. 

\section{A distorted rotating black ring}
\subsection{Distorted Minkowski Seed solution}
We begin by considering a distorted five-dimensional Minkowski spacetime as a static seed solution. The distortions preserve Ricci-flatness, although the resulting seed will not be flat. To obtain the solutions with
sufficient variety, however, we add a freedom of one parameter
to the seed metric. Consider Minkowski spacetime 
\begin{equation}
    ds^2 = -(dx_0)^2 + (dr^2 + r^2 d\phi^2) + (d\chi^2 + \chi^2 d\psi^2)  
\end{equation} expressed in the form $\mathbb{R} \times \mathbb{R}^2 \times \mathbb{R}^2$ with $(r,\phi)$ and $(\chi, \psi)$ standard polar coordinates on each Euclidean plane. By introducing cylindrical-type coordinates
\begin{equation}
    \rho = r \chi, \qquad z = \dfrac{1}{2} (\chi^2 - r^2) - \lambda \sigma,
\end{equation} where $ \lambda $ is an arbitrary real constant, the metric takes the form
\begin{equation}\label{UndistortedMin}
		\begin{aligned}
	ds^{2}&=-(dx_{0})^{2}+\left(\sqrt{\rho^{2}+(z+\lambda \sigma)^{2}}-(z+\lambda \sigma)\right)(d\phi)^{2}+\left(\sqrt{\rho^{2}+(z+\lambda \sigma)^{2}}+z+\lambda \sigma\right)(d\psi)^{2} \\
	&+\dfrac{1}{2\sqrt{\rho^{2}+(z+\sigma \lambda)^{2}}}\left(d\rho^{2}+dz^{2}\right) \\
	&=-(dx_{0})^{2}+\sigma \left(\sqrt{(x^{2}-1)(1-y^{2})+(xy+\lambda)^{2}}-(xy+\lambda)\right)(d\phi)^{2}\\
	&+\sigma\
\left(\sqrt{(x^{2}-1)(1-y^{2})+(xy+\lambda)^{2}}+(xy+\lambda)\right)(d\psi)^{2}	\\
&+\dfrac{\sigma(x^{2}-y^{2})}{2\sqrt{(x^{2}-1)(1-y^{2})+(xy+\lambda)^{2}}}\left(\dfrac{dx^{2}}{x^{2}-1}+\dfrac{dy^{2}}{1-y^{2}}\right)	
\end{aligned}
\end{equation} where we have used the prolate spheroidal coodinates $(x,y)$ introduced above in the second equality. 
 To construct the desired solutions, we will need to impose $\lambda > 1$.
By comparing this metric with \eqref{WPstatic}, we can read off the corresponding values of $ S $ and $ T $: 
\begin{equation}
	\begin{aligned}
			S_{(0)}=T_{(0)}&=\frac{1}{2}\ln\left[\sqrt{\rho^{2}+(z+\lambda \sigma)^{2}}+(z+\lambda \sigma)\right]\\
			&=\frac{1}{2}\ln\left[\sigma\left(\sqrt{(x^{2}-1)(1-y^{2})+(xy+\lambda)^{2}}+(xy+\lambda)\right)\right].
	\end{aligned}
\end{equation}
We now rewrite the five-dimensional Minkowski spacetime metric in prolate spheroidal coordinates in the general form (see, for comparison \cite{Tavayef2022})
\begin{equation}\label{undistortedstatic}
	ds^{2}=-(dt)^{2}+e^{-2(\widetilde{U})}(d\phi)^{2}+e^{-2(\widetilde{W})}(d\psi)^{2}+\frac{e^{2\widetilde{V}}(x^2-y^2)}{(x^2-1)(1-y^2)} \left[\frac{dx^2}{x^2-1} + \frac{dy^2}{1-y^2}\right]	.
\end{equation}
By comparing eq. (\ref{UndistortedMin}) with the above metric, we reasd off
\begin{equation}\label{Wt}
	\widetilde{W}=\left(-\frac{1}{2}\right) \ln 
	\left(\sigma\sqrt{(x^{2}-1)(1-y^{2})+(xy+\lambda)^{2}}+\sigma(xy+\lambda)\right) = -S_{(0)},
\end{equation}
\begin{equation}\label{Ut}
	\widetilde{U}=\left(-\frac{1}{2}\right) \ln \left( \sigma\sqrt{(x^{2}-1)(1-y^{2})+(xy+\lambda)^{2}}- \sigma(xy+\lambda)\right)
\end{equation}
\begin{equation}\label{Vt}
\widetilde{V} = \left(\frac{1}{2}\right) \ln \left(\frac{\sigma (x^2-1)(1-y^2)}{2\sqrt{(x^{2}-1)(1-y^{2})+(xy+\lambda)^2}}\right).
\end{equation}

The functions $ \widetilde{U} $ and $ \widetilde{W} $ are solutions to the Laplace
equation in 3D flat space, and in Weyl coordinates their
sum satisfies $ -\widetilde{U} -\widetilde{W}=\ln \rho  $, as required for a 5D Weyl
solution \cite{EmparanReall}. 
By comparing the metric \eqref{WPstatic} with \eqref{undistortedstatic} for the undistorted seed
we have
\begin{equation}
\exp\left(2\tilde{\gamma_{0}}-\tilde{S_{0}}-\tilde{T_{0}}\right) = \frac{\exp({2\tilde{V}})}{\rho^{2}}.
\end{equation}
By substituting $S_{0}=T_{0}=-\widetilde{W}$ we get
\begin{equation}
e^{\widetilde{\gamma_{0}}}=\frac{e^{(\widetilde{V}-\widetilde{W})}}{\rho}.
\end{equation}
By replacing $\widetilde{U}  $ and $\widetilde{W}$ with appropriately modified axisymmetric solutions of Laplace's equation $\widetilde{U} + \widehat{U}, \widetilde{W} + \widehat{W}$ and calculating the associated metric function $\gamma $, we can produce an asymptotically non-flat, Ricci-flat Weyl solution that does not contain any horizons. It can be referred to as a distorted Minkowski spacetime, analogous to distorted black hole solutions (e.g., the distorted black hole-bubble solutions of \cite{Tavayef2022}). 
The functions $ \widehat{U} $ and $ \widehat{W} $ characterize the external gravitational sources that produce these distortions. The most general form of them is given by the expressions
\begin{equation}\label{Uhat}
	\widehat{U}=\sum_{n= 1}^{\infty} a_{n}R^{n}P_{n}\left(\dfrac{xy}{R}\right) , \quad 	\widehat{W}=\sum_{n=1}^{\infty} b_{n}R^{n}P_{n}\left(\dfrac{xy}{R}\right), 
\end{equation}
where
\begin{equation}
	R=\dfrac{\sqrt{\rho^{2}+z^{2}}}{\sigma}=\sqrt{x^{2}+y^{2}-1}.
\end{equation} Note that we have only considered `external' sources whose contributions vanish as $R \to 0$ but not on the axis and horizon ($\rho \to 0$). This will allow us to remove conical singularities while the horizon geometry will be distorted. Here, $P_{n}(xy/R)$ represents the $n-$th Legendre polynomials of the first kind, $ a_{n} $ and $b_{n}$ are real constants, and $n$ is a natural number.

\subsection{Construction of the local solution}
For the general static and axisymmetric external gravitational fields, the metric of the distorted Minkowski space-time has the form
\begin{equation}\label{distortedstatic}
\begin{aligned}
	ds^{2}&=-e^{2(\widehat{U}+\widehat{W})}(dx_{0})^{2}+e^{-2(\widetilde{U}+\widehat{U})}(d\phi)^{2}+e^{-2(\widetilde{W}+\widehat{W})}(d\psi)^{2} \\ & +\frac{e^{2(\widetilde{V}+\widehat{V}+\widehat{U}+\widehat{W})}(x^2-y^2)}{(x^2-1)(1-y^2)} \left(\dfrac{dx^{2}}{x^{2}-1}+\dfrac{dy^{2}}{1-y^{2}}\right). 
    \end{aligned}
\end{equation}
We will utilize the twofold Bäcklund transformation on the distorted Minkowski spacetime \eqref{distortedstatic} as the seed. This transformation can be applied with respect to either of the spacelike Killing fields, resulting in a solution for the given angular direction. 
By choosing the Killing field $ \partial /\partial \phi $ as an axis of rotation and comparing the metric of the distorted Minkowski spacetime \eqref{distortedstatic} with the general expression \eqref{WPstatic}, we have the identifications
\begin{equation}\label{T0S0}
	T_{(0)}=-\widetilde{W}-\widehat{W},\quad S_{(0)}=-\widetilde{W}+\widehat{W}+2\widehat{U}
\end{equation} Comparing the metric \ref{WPstatic} with \ref{undistortedstatic} for the distorted seed, we have
\begin{equation}
e^{\left(2\gamma_{0}-S_{0}-T_{0}\right) }= e^{2(V+U+W)}.
\end{equation} By substituting $S_{0}$ and $T_{0}$ we get
\begin{equation}
e^{\left(2\widetilde{\gamma_{0}}+2\widehat{\gamma_{0}}+\widetilde{W}-\widehat{W}-2\widehat{U}+\widetilde{W}+\widehat{W}\right) }= \frac{e^{2\tilde{V}}}{\rho^{2}} e^{2(\widehat{V}+\widehat{U}+\widehat{W})},
\end{equation} 
where we have written $\gamma_0 = \widetilde{\gamma_{0}} + \widehat{\gamma_{0}}$. Rearranging the previous expression gives
\begin{equation}
e^{\left(\widetilde{\gamma_{0}}+\widehat{\gamma_{0}}\right)} = \frac{e^{(\widetilde{V}-\widetilde{W})}}{\rho} e^{(\widehat{V}+\widehat{W}+2\widehat{U})}.
\end{equation} Next, if we substitute $ S_{(0)} $ in \eqref{diffa-gl}
\begin{align}\nonumber
	(\ln a)_{,x}&=\dfrac{1}{x-y}\left[(xy-1)\left(-\widetilde{W}_{,x}+(\widehat{W}+2\widehat{U})_{,x}\right)+(1-y^{2})\left(-\widetilde{W}_{,y}+(\widehat{W}+2\widehat{U})_{,y}\right)\right],\quad \\ 
	(\ln a)_{,y}&=\dfrac{1}{x-y}[-(x^{2}-1)\left(-\widetilde{W}_{,x}+(\widehat{W}+2\widehat{U})_{,x}\right)+(xy-1)\left(-\widetilde{W}_{,y}+(\widehat{W}+2\widehat{U})_{,y}\right)].
	\nonumber
\end{align} This can organize these into a part from the background solution and the distortions (i.e. quantities marked with a tilde and a hat, respectively):
\begin{align}\nonumber
	(\ln a)_{,x}&=\left(\ln \tilde{a}\right)_{,x}+\dfrac{1}{x-y}\left[(xy-1)\left(\widehat{W}+2\widehat{U}\right)_{,x}+(1-y^{2})\left(\widehat{W}+2\widehat{U}\right)_{,y}\right]\\ \nonumber
	&=\left(\ln \tilde{a}\right)_{,x}+\left(\ln \hat{a}\right)_{,x}=\left(\ln (\tilde{a}\hat{a})\right)_{,x},\nonumber
\end{align}
\begin{align*}
		(\ln a)_{,y}&=\left(\ln \tilde{a}\right)_{,y}+\dfrac{1}{x-y}\left[-(x^{2}-1)\left(\widehat{W}+2\widehat{U}\right)_{,x}+(xy-1)\left(\widehat{W}+2\widehat{U}\right)_{,y}\right]\\
		&=\left(\ln \tilde{a}\right)_{,y}+\left(\ln \hat{a}\right)_{,y}=\left(\ln (\tilde{a}\hat{a})\right)_{,y}.
	\nonumber
\end{align*}
Also, by substituting $ S_{(0)} $ in eqs.\ref{diffb-gl}
\begin{eqnarray*}
	(\ln b)_{,x}=-\dfrac{1}{x+y}\left[(xy+1)\left(-\widetilde{W}_{,x}+(\widehat{W}+2\widehat{U})_{,x}\right)+(1-y^{2})\left(-\widetilde{W}_{,y}+(\widehat{W}+2\widehat{U})_{,y}\right)\right],\\ \nonumber
		(\ln b)_{,y}=-\dfrac{1}{x+y}\left[-(x^{2}-1)\left(-\widetilde{W}_{,x}+(\widehat{W}+2\widehat{U})_{,x}\right)+(xy+1)\left(-\widetilde{W}_{,y}+(\widehat{W}+2\widehat{U})_{,y}\right)\right]
	\nonumber
\end{eqnarray*}
\begin{eqnarray*}
	(\ln b)_{,x}&&=(\ln \tilde{b})_{,x}-\dfrac{1}{x+y}\left[(xy+1)\left(\widehat{W}+2\widehat{U}\right)_{,x}+(1-y^{2})\left(\widehat{W}+2\widehat{U}\right)_{,y}\right]\\
	&&=\left(\ln \tilde{b}\right)_{,x}+\left(\ln \hat{b}\right)_{,x}=\left(\ln (\tilde{b}\hat{b})\right)_{,x},\\ \quad
	(\ln b)_{,y}&&=(\ln \tilde{b})_{,y}-\dfrac{1}{x+y}\left[-(x^{2}-1)\left(\widehat{W}+2\widehat{U}\right)_{,x}+(xy+1)\left(\widehat{W}+2\widehat{U}\right)_{,y}\right]\\
	&&=\left(\ln \tilde{b}\right)_{,y}+\left(\ln \hat{b}\right)_{,y}
	=\left(\ln (\tilde{b}\hat{b})\right)_{,y},\quad \\  
\end{eqnarray*}
The first-order equations for $ a = a (x,y) $ and $b = b(x,y)$ of a similar form were solved previously in \cite{breton1998} in the construction of distorted rotating Kerr solutions. Modifying their approach appropriately, we find that
\begin{equation}\label{a}
a=\tilde{a}\left(-\kappa \exp{\sum_{n= 1}^{\infty} (b_{n}+2a_{n})(x-y)\sum_{l= 0}^{n-1} R^{l}P_{l}\left(\dfrac{xy}{R}\right)}\right)
\end{equation}
\begin{equation}\label{b}
	b=\tilde{b}\left(-\kappa \exp{\sum_{n= 1}^{\infty} (b_{n}+2a_{n})(x+y)\sum_{l= 0}^{n-1} (-1)^{(n-l)}R^{l}P_{l}\left(\dfrac{xy}{R}\right)}\right)
\end{equation}
where $\tilde{a} $ and $ \tilde{b} $ are given by (c.f.\cite{iguchi2006})
\begin{equation}
\tilde{a}=\alpha \dfrac{(x-y+1+\lambda)+\sqrt{x^{2}+y^{2}+2\lambda xy+(\lambda^{2}-1)}}{2\left[(xy+\lambda)+\sqrt{x^{2}+y^{2}+2\lambda xy+(\lambda^{2}-1)}\right]^{1/2}}
\end{equation}
\begin{equation}
	\tilde{b}=\beta \dfrac{2\left[(xy+\lambda)+\sqrt{x^{2}+y^{2}+2\lambda xy+(\lambda^{2}-1)}\right]^{1/2}}{(x+y-1+\lambda)+\sqrt{x^{2}+y^{2}+2\lambda xy+(\lambda^{2}-1)}}.
\end{equation} We will absorb the arbitrary constant $\kappa$ in the parameters $\alpha, \beta$. 
Next, we must integrate to find $ \gamma^{\prime} $ in the metric \eqref{static2}. By substituting \eqref{T0S0} into \eqref{Tp,Sp} one finds
\begin{align}
S^{\prime}&=2U_{0}^{(BH)}-\widetilde{W}+\widehat{W}+2\widehat{U}, \\  
T^{\prime}&=-\widetilde{W}-\widehat{W}.
\end{align}
By substituting them in eq.(\ref{diffr-Gprime}) we have
\begin{eqnarray}\nonumber
\partial_{\rho}\gamma^{\prime}&=&\frac{\rho}{4} \left[(\partial_{\rho}(S^{\prime}))^{2}-(\partial_{z}S^{\prime})^{2}\right]+\frac{3\rho}{4} \left[(\partial_{\rho}T^{\prime})^{2}-(\partial_{z}T^{\prime})^{2}\right] \\ \nonumber
	&=&\frac{\rho}{4} \left[\left(\partial_{\rho}(2U_{0}^{(BH)}-\widetilde{W})+\partial_{\rho}(\widehat{W}+2\widehat{U})\right)^{2}-\left(\partial_{z}(2U_{0}^{(BH)}-\widetilde{W})+\partial_{z}(\widehat{W}+2\widehat{U})\right)^{2}\right]+\\ \nonumber
	&&\frac{3\rho}{4} \left[(\partial_{\rho}(-\widetilde{W}-\widehat{W}))^{2}-(\partial_{z}(-\widetilde{W}-\widehat{W}))^{2}\right]\\ \nonumber &=&\dfrac{\rho}{4}\left[(\partial_{\rho}(2U_{0}^{(BH)}-\widetilde{W}))^{2}-(\partial_{z}(2U_{0}^{(BH)}-\widetilde{W}))^{2}\right]+\dfrac{3\rho}{4}\left[(\partial_{\rho}\widetilde{W})^{2}-(\partial_{z}\widetilde{W})^{2}\right]\\ \nonumber
	&&+\dfrac{\rho}{4}\left[(\partial_{\rho}(\widehat{W}+2\widehat{U} ))^{2}-(\partial_{z}(\widehat{W}+2\widehat{U} ))^{2}\right]+\dfrac{3\rho}{4}\left[(\partial_{\rho}\widehat{W})^{2}-(\partial_{z}\widehat{W})^{2}\right]\\ \nonumber
	&&+\dfrac{\rho}{4}\left[(\partial_{\rho}(2U_{0}^{(BH)}-\widetilde{W}))(\partial_{\rho}\widehat{W}+2\widehat{U})-2(\partial_{z}(2U_{0}^{(BH)}-\widetilde{W} ))(\partial_{z}(\widehat{W}+2\widehat{U}))\right]\\ \nonumber
	&&+\dfrac{3\rho}{4}\left[2(\partial_{\rho}\widehat{W})(\partial_{\rho}\widehat{W})-(2(\partial_{z}\widetilde{W})(\partial_{z}\widehat{W}))\right]=\partial_{\rho}\widetilde{\gamma^{\prime}}+\partial_{\rho}\widehat{\gamma^{\prime}_{1}}+\partial_{\rho}\widehat{\gamma^{\prime}_{2}}
\end{eqnarray}
Similarly, by substituting them in \eqref{diffz-Gprime} we obtain
\begin{eqnarray*}
\partial_{z}\gamma^{\prime}&&=\dfrac{\rho}{2}\left[\partial_{\rho}(2U_{0}^{(BH)}-\widetilde{W})\partial_{z}(2U_{0}^{(BH)}-\widetilde{W})\right]+\dfrac{3\rho}{2}(\partial_{\rho}\widetilde{W}\partial_{z}\widetilde{W})\\ 
	&&+\dfrac{\rho}{2}\left[\partial_{\rho}(\widehat{W}+2\widehat{U})\partial_{z}(\widehat{W}+2\widehat{U})\right]+\dfrac{3\rho}{2}(\partial_{\rho}\widehat{W}\partial_{z}\widehat{W})\\ 
	&&+\dfrac{\rho}{2}\left[\partial_{\rho}(2U_{0}^{(BH)}-\widetilde{W})\partial_{z}(\widehat{W}+2\widehat{U})+\partial_{z}(2U_{0}^{(BH)}-\widetilde{W})\partial_{\rho}(\widehat{W}+2\widehat{U})\right]\\
	&&+\dfrac{3\rho}{2}(\partial_{\rho}\widetilde{W}\partial_{z}\widehat{W}+\partial_{z}\widetilde{W}\partial_{\rho}\widehat{W})=\partial_{z}\widetilde{\gamma^{\prime}}+\partial_{z}\widehat{\gamma^{\prime}}_{1}+\partial_{z}\widehat{\gamma^{\prime}}_{2}
\end{eqnarray*} where $ \widehat{\gamma^{\prime}}_{1}+	\widehat{\gamma^{\prime}}_{2} =	\widehat{\gamma^{\prime}}$ comprises the total distorted part of $  \gamma^{\prime}$ and $\widetilde{\gamma^{\prime}}$ is understood to be determined in terms of background (undistorted) quantities. These have been treated previously with the result
\begin{eqnarray}
		\widetilde{\gamma^{\prime}}&&=\dfrac{1}{4}\ln \left(\sqrt{\rho^{2}+(z-\sigma)^{2}}+z-\sigma\right)-\dfrac{1}{4}\ln \left(\sqrt{\rho^{2}+(z+\sigma)^{2}}+z+\sigma\right)\\ \nonumber
		&&+\dfrac{1}{2}\ln \left(\sqrt{\rho^{2}+(z+\lambda\sigma)^{2}}+z+\lambda\sigma\right)
		-\dfrac{1}{4}\ln \left(\sqrt{\rho^{2}+(z+\sigma)^{2}}\right)-\dfrac{1}{4}\ln \left(\sqrt{\rho^{2}+(z-\sigma)^{2}}\right) \\ \nonumber
		&& -\dfrac{1}{4}\ln \left(\sqrt{\rho^{2}+(z+\lambda\sigma)^{2}}\right) +\dfrac{1}{2}\ln \left(\sqrt{\rho^{2}+(z-\sigma)^{2}}\sqrt{\rho^{2}+(z+\sigma)^{2}}+(z-\sigma)(z+\sigma)+\rho^{2}\right)\\ \nonumber
	&&-\dfrac{1}{4}\ln \left(\sqrt{\rho^{2}+(z-\sigma)^{2}}\sqrt{\rho^{2}+(z+\lambda\sigma)^{2}}+(z-\sigma)(z+\lambda\sigma)+\rho^{2}\right)\\ \nonumber
	&&+\dfrac{1}{4}\ln \left(\sqrt{\rho^{2}+(z+\sigma)^{2}}\sqrt{\rho^{2}+(z+\lambda\sigma)^{2}}+(z+\sigma)(z+\lambda\sigma)+\rho^{2}\right)-\dfrac{3}{4}\ln 2. \nonumber
\end{eqnarray} The distorted quantities appearing in this decomposition satisfy the following.
\begin{eqnarray}\label{Diffgamma1}
	\partial_{\rho}\widehat{\gamma^{\prime}}_{1}&&=\dfrac{\rho}{4}\left[(\partial_{\rho}(\widehat{W}+2\widehat{U} ))^{2}-(\partial_{z}(\widehat{W}+2\widehat{U} ))^{2}\right]+\dfrac{3\rho}{4}\left[(\partial_{\rho}\widehat{W})^{2}-(\partial_{z}\widehat{W})^{2}\right]\\
		\partial_{z}\widehat{\gamma^{\prime}}_{1}&&=\dfrac{\rho}{2}\left[\partial_{\rho}(\widehat{W}+2\widehat{U})\partial_{z}(\widehat{W}+2\widehat{U})\right]+\dfrac{3\rho}{2}(\partial_{\rho}\widehat{W}\partial_{z}\widehat{W})\nonumber	
\end{eqnarray}
and lastly, writing $\widehat{\gamma^{\prime}}_{2}=	\widehat{\gamma^{\prime}}_{21}+	\widehat{\gamma^{\prime}}_{22} $, we have 
\begin{eqnarray}
\partial_{\rho}\widehat{\gamma^{\prime}}_{21}&&=\rho\left[\partial_{\rho}(U_{0}^{(BH)})\partial_{\rho}(\widehat{W}+2\widehat{U})-\partial_{z}(U_{0}^{(BH)})\partial_{z},(\widehat{W}+2\widehat{U})\right] ,\\ \nonumber
			\partial_{z}\widehat{\gamma^{\prime}}_{21}&&=\rho\left[\partial_{\rho}(U_{0}^{(BH)})\partial_{z}(\widehat{W}+2\widehat{U})+\partial_{z}(U_{0}^{(BH)})\partial_{\rho}(\widehat{W}+2\widehat{U})\right],
\end{eqnarray} and 
\begin{eqnarray}\label{gamma22}
	\partial_{\rho}	\widehat{\gamma^{\prime}}_{22}&&=\rho\left[\widetilde{W}_{,\rho}\left(\widehat{W}-\widehat{U}\right)_{,\rho}-\widetilde{W}_{,z}\left(\widehat{W}-\widehat{U}\right)_{,z}\right],\\
	\partial_{z}	\widehat{\gamma^{\prime}}_{22}&&=\rho\left[\widetilde{W}_{,\rho}\left(\widehat{W}-\widehat{U}\right)_{,z}+\widetilde{W}_{,z}\left(\widehat{W}-\widehat{U}\right)_{,\rho}\right].\nonumber
\end{eqnarray} These equations can be solved following a similar approach as in \cite{breton1998}. The results are
\begin{equation}
		\widehat{\gamma^{\prime}}_{1}=\dfrac{1}{4}\sum_{n,k= 1}^{\infty} \dfrac{nk\left((b_{n}+2a_{n})(b_{k}+2a_{k})+3b_{n}b_{k}\right)R^{n+k}}{n+k}\left(P_{n}P_{k}-P_{n-1}P_{k-1}\right)
\end{equation}  and 
\begin{equation}
	\widehat{\gamma^{\prime}}_{21}=\sum_{n= 1}^{\infty} \dfrac{b_{n}+2a_{n}}{2}\sum_{l= 0}^{n-1}\left((-1)^{n-l+1}(x+y)-x+y\right)R^{l}P_{l}\left(\dfrac{xy}{R}\right).
\end{equation} We can solve these rather complicated equations by a series of further decompositions, which yield equations of the simplified form: 
\begin{eqnarray}\label{gamma230:1}
	\gamma_{,\rho} = \rho \left(\psi_{,\rho}^2 - \psi_{,z}^2\right),\qquad \gamma_{,z} = 2 \rho \, \psi_{,\rho} \psi_{,z}. \nonumber
\end{eqnarray} where $ \psi $ is of the form
\begin{eqnarray}
	\psi &&= \frac{1}{2} \ln \frac{x - 1}{x + 1} + \sum_{n=1}^{\infty} a_n R^n P_n \left(\frac{xy}{R}\right)=:U_{0}^{(BH)}+\widehat{Y} \\ 
	R &&\equiv \sqrt{x^2 + y^2 - 1}. \nonumber
\end{eqnarray}
The solution can be obtained by straightforward manipulations from \cite[Eq 3]{breton1998}
\begin{eqnarray}
	\gamma&& = \frac{1}{2} \ln \frac{x^2 - 1}{x^2 - y^2}
	+ \sum_{n,p=1}^{\infty} \frac{n p \, a_n a_p \, R^{n+p}}{n + p} \left(P_n P_p - P_{n-1} P_{p-1}\right)\\ \nonumber
	&&+ \sum_{n=1}^{\infty} a_n \sum_{l=0}^{n-1} \left[(-1)^{n-l+1}(x + y) - x + y\right] R^l P_l,
\end{eqnarray}
Consider integrating equations of the form
\begin{eqnarray*}
	\gamma_{,\rho} &&= \rho \left(((U_{0}^{(BH)})_{,\rho}+\widehat{Y}_{,\rho})^2 -((U_{0}^{(BH)})_{,z}+\widehat{Y}_{,z})^2 \right)\\
	&&=\rho\left[\partial_{\rho}(U_{0}^{(BH)})^{2}-\partial_{z}(U_{0}^{(BH)})^{2}\right]+\rho\left[\partial_{\rho}(\widehat{Y})^{2}-\partial_{z}(\widehat{Y})^{2}\right]\\
	&&+2\rho\left[\partial_{\rho}(U_{0}^{(BH)})\partial_{\rho}(\widehat{Y})-\partial_{z}(U_{0}^{(BH)})\partial_{z}(\widehat{Y})\right]=\partial_{\rho}\gamma^{\prime}_{BH}+\partial_{\rho}\gamma^{\prime}_{\widehat{Y}}+\partial_{\rho}\gamma^{\prime}_{c}
\end{eqnarray*} and
\begin{eqnarray*}
		\gamma_{,z} &&= 2 \rho \,\left[(U_{0}^{(BH)})_{,\rho}+\widehat{Y}_{,\rho}\right] \left[(U_{0}^{(BH)})_{,z}+\widehat{Y}_{,z}\right]\\
		&&=(U_{0}^{(BH)})_{,\rho}(U_{0}^{(BH)})_{,z}+(\widehat{Y})_{,\rho}(\widehat{Y})_{,z}+(U_{0}^{(BH)})_{,\rho}(\widehat{Y})_{,z}+(U_{0}^{(BH)})_{,z}(\widehat{Y})_{,\rho}\\
		&&=\partial_{z}\gamma^{\prime}_{BH}+\partial_{z}\gamma^{\prime}_{\widehat{Y}}+\partial_{z}\gamma^{\prime}_{c}.
\end{eqnarray*}
Comparing with \eqref{gamma230:1}, we see the solution of the first term is
\begin{equation*}
\gamma^{\prime}_{BH}=\frac{1}{2} \ln \frac{x^2 - 1}{x^2 - y^2}.
\end{equation*}
The solution for the second term is
\begin{equation}
	\gamma^{\prime}_{\widehat{Y}}=\sum_{n,p=1}^{\infty} \frac{n p \, a_n a_p \, R^{n+p}}{n + p} \left(P_n P_p - P_{n-1} P_{p-1}\right) \nonumber
\end{equation}
and the solution of the third term is
\begin{equation}
	\gamma^{\prime}_{c}= \sum_{n=1}^{\infty} a_n \sum_{l=0}^{n-1} \left[(-1)^{n-l+1}(x + y) - x + y\right] R^l P_l . \nonumber
\end{equation} For example, for the equation determining $\widehat{\gamma^{\prime}}_{21}$, we have $ \psi=U_{0}^{(BH)}+\widehat{F} $ where $ \widehat{F}=\widehat{W}+2\widehat{U}=\sum_{n=1}^{\infty} (b_n+2a_n) R^n P_n \left(xy/R\right) $.  Similarly, comparing to the above equations and solutions, we can write
\begin{eqnarray}
	\partial_{\rho}\widehat{\gamma^{\prime}}_{1}&&=\dfrac{1}{4}\partial_{\rho}\widehat{\gamma^{\prime}}_{\widehat{F}}+\dfrac{3}{4}\partial_{\rho}\widehat{\gamma^{\prime}}_{\widehat{W}}\\
	\partial_{z}\widehat{\gamma^{\prime}}_{1}&&=\dfrac{1}{4}\partial_{z}\widehat{\gamma^{\prime}}_{\widehat{F}}+\dfrac{3}{4}\partial_{z}\widehat{\gamma^{\prime}}_{\widehat{W}}
\end{eqnarray} This leads to the above solutions for $\widetilde{\gamma_1^\prime}$ and $\widetilde{\gamma_{21}^\prime}$.
Finally, by substituting \eqref{Wt} and \eqref{Uhat} into the above equations for $\widehat{\gamma^{\prime}}_{22}$ \eqref{gamma22}, although we have not been able to find a unified formula for general $n$, we can explicitly obtain expressions 
for $\widehat{\gamma^{\prime}}_{22}$ for the multiple moments $ n=1,2,3,... $,
\begin{align}
		\widehat{\gamma^{\prime}}_{22} &=\dfrac{b_{1}-a_{1}}{2\sigma}\left(\sqrt{\rho^{2}+(z+\lambda \sigma)^2}-z\right), \quad n=1 \\
			\widehat{\gamma^{\prime}}_{22}&=\dfrac{b_{2}-a_{2}}{2\sigma^{2}}\left(\sqrt{\rho^{2}+(z+\lambda\sigma)^2}(z-\lambda \sigma)-z^{2}+\dfrac{\rho^{2}}{2}\right),\quad n=2 \\
				\widehat{\gamma^{\prime}}_{22}&=\dfrac{b_{3}-a_{3}}{2\sigma^{3}}\left(\sqrt{\rho^{2}+(z+\lambda\sigma)^2}((z-\lambda \sigma)^{2}-\dfrac{\rho^{2}}{2}+\lambda\sigma z)-z(z^{2}-\dfrac{3\rho^{2}}{2})\right),\quad n=3		\\
         \widehat{\gamma^{\prime}}_{22} &= \frac{(a_4 - b_4)}{2\sigma^4} 
\left(
- z^3 + \lambda \sigma z^2 
+ \left(-\lambda^2 \sigma^2 + \tfrac{3}{2}\rho^2\right) z 
+ \lambda^3 \sigma^3 - \tfrac{1}{2}\rho^2\sigma\lambda
\right)
\sqrt{\rho^2 + (z+\lambda\sigma)^2}\\ \nonumber
&+ \frac{(a_4 - b_4)}{2\sigma^4} \left(z^4 - 3\rho^2 z^2 + \tfrac{3}{8}\rho^4\right), \quad n=4 
\end{align} and for $n=5$,
\begin{eqnarray}
   \widehat{\gamma^{\prime}}_{22} &= \frac{(a_5 - b_5)}{2\sigma^5} 
\left( -z^4 + \lambda \sigma z^3 + \left( -\lambda^2 \sigma^2 + 3\rho^2 \right) z^2 
+ \left( \lambda^3 \sigma^3 - \frac{3}{2} \rho^2 \sigma \lambda \right) z \right)\sqrt{\rho^{2}+(z+\lambda \sigma)^2}\\ \nonumber
&+ \frac{(a_5 - b_5)}{2\sigma^5}\left(\left(- \lambda^4 \sigma^4 +\frac{\lambda^2 \sigma^2 \rho^2}{2} - \frac{3\rho^4}{8} \right)
\sqrt{\rho^{2}+(z+\lambda \sigma)^2}
+ z \left( z^4 - 5\rho^2 z^2 + \frac{15}{8} \rho^4 \right) 
\right).
\end{eqnarray} We have thus determined $\gamma^{\prime}_{total}=	\widetilde{\gamma^{\prime}}+\widehat{\gamma^{\prime}}_{1}+	\widehat{\gamma^{\prime}}_{2}=\widetilde{\gamma'}+\widehat{\gamma'} $.
We can then explicitly obtain the local metric of the distorted solution: 
\begin{equation}\label{metricxy}
\begin{aligned}
		ds^{2}&=-\dfrac{\hat{A}}{\hat{B}}e^{2(\widehat{W}+\widehat{U})}\left[dx_{0}-\left(2\sigma e^{\widetilde{W}-\widehat{W}-2\widehat{U}}\dfrac{\hat{C}}{\hat{A}}+\hat{c}_{1}\right)d\phi\right]^{2}\\
		&+\sigma^{2}(x^{2}-1)(1-y^{2})\dfrac{\hat{B}}{\hat{A}}e^{2(\widetilde{W}-\widehat{U})}d\phi^{2}
		+e^{-2(\widetilde{W}-\widehat{W})}d\psi^{2}\\ 
		&+\left(\dfrac{\hat{c}_{2}(x^{2}-y^{2})\hat{B}e^{2(\widetilde{\gamma}^{\prime}+\widetilde{W})}e^{2(\hat{\gamma^{\prime}}-\widehat{U})}}{x^{2}-1}\right)\left(\dfrac{dx^2}{x^{2}-1}+\dfrac{dy^{2}}{1-y^{2}}\right),
\end{aligned}
\end{equation}
where $\hat{A},\hat{B},\hat{C}$ are given by \eqref{A}, \eqref{B}, \eqref{C} respectively with $a,b$ now given by \eqref{a}, \eqref{b}. The prolate-spheroidal coordinates, $x$ and $y$, can be written explicitly in terms of $\rho$ and $z$ by inverting \eqref{rhoz}. In the canonical coordinates $\rho$ and $z$ we have
\begin{equation}
    x=\dfrac{\sqrt{\rho^2+(z+\sigma)^2}+\sqrt{\rho^2+(z-\sigma)^2}}{2\sigma},\quad  y=\dfrac{\sqrt{\rho^2+(z+\sigma)^2}-\sqrt{\rho^2+(z-\sigma)^2}}{2\sigma}.
\end{equation}
If we substitute them in the metric \ref{metricxy}, we obtain
\begin{eqnarray}\label{metricrz1}
	ds^{2}&=&-\dfrac{\hat{A}}{\hat{B}}e^{2(\widehat{W}+\widehat{U})}\left(dx_{0}-(2\sigma e^{\widetilde{W}-\widehat{W}-2\widehat{U}}\dfrac{\hat{C}}{\hat{A}}+\hat{c}_{1})d\phi\right)^{2}+\rho^{2}\dfrac{\hat{B}}{\hat{A}}e^{2(\widetilde{W}-\widehat{U})}d\phi^{2}\\\nonumber
    &&+e^{-2(\widetilde{W}-\widehat{W})}d\psi^{2}
+\dfrac{2\hat{c}_{2}\sigma^2\hat{B}e^{2(\widetilde{\gamma}^{\prime}+\widetilde{W})}e^{2(\hat{\gamma^{\prime}}-\hat{U})}}{z^2 + \rho^2 - \sigma^2 + \sqrt{\sigma^4 + (2\rho^2 - 2z^2)\sigma^2 + (\rho^2 + z^2)^2}}\left(d\rho^2+dz^2\right).
\end{eqnarray} In summary, \eqref{metricrz1} constitutes a stationary, biaxisymmetric solution for the vacuum equations parameterized by $\{\sigma,\lambda,a_n, b_n \}$ and the integration constants $\{ \hat{c}_1, \hat{c}_2\}$. 

\subsection{Rod Data Set} The topology of the spacetime is determined by the zero set of the determinant of the metric restricted to the Killing fields. For the distorted metric \eqref{metricxy} this is given by 
\begin{equation}\label{hatrho}
    \widehat{\rho} = \sigma \sqrt{(x^2-1)(1-y^2)}
\end{equation} which is the same as in the undistorted case, i.e. $\widehat{\rho} = \rho$. Hence, the distortions do not affect the rod structure, although they do change the behaviour of the metric near the zero set, which allows us to remove conical singularities and modify the geometry of the horizon, as well as possibly altering the rod vectors which define which combination of Killing fields either vanishes or goes null on the zero set of $\rho$. 

We will choose the range of $(x,y)$ so that $x \geq 1, -1 \leq y \leq 1$. It is easily seen that $\rho =0$ when $x = 1$ and $y = \pm 1$. Observe that $e^{-2\widetilde{W}}$ is a smooth non-zero function of $x$ on $y = 1$, but that
\begin{equation}
e^{-2\widetilde{W}}\vert_{y = -1} = \begin{cases} 0 & x > \lambda >1 \\ 2\sigma(\lambda - x) &1 <  x < \lambda \end{cases} .
\end{equation} Furthermore, since $\lambda > 1$, we observe that on the set $x = 1$, $e^{-2\widetilde{W}} = 2\sigma(y + \lambda) > 0$. Finally, as observed above, the distortion fields $\widehat{U}, \widehat{W}$ are smooth, bounded functions and do not introduce new rods. 

\subsubsection{Undistorted solution}\label{undistorted}
 First, we consider the undistorted solution. It is easily checked that $\hat{B}$ is smooth, bounded, and positive everywhere along $x = 1, y = \pm 1$. $\hat{A}$ is smooth and bounded along $x=1$ and $y = \pm 1$ and vanishes to first order on approach to the rod points. Finally, on the rods $x = 1$ and  $y = 1$, $\hat{C} \neq 0$ (in particular, $\hat{C} = O(1-y^2)$ and $\hat{C} = O(1-x^2)$ respectively). Similarly, along $y = -1$, $\hat{C} = O(1-x^2)$ for $1 < x < \lambda$ and is non-vanishing. However, $\hat{C} =0$ along $y = -1, x > \lambda$.  This gives the following rod structure. There are two semi-infinite rods: (1) $y = -1, x > \lambda > 1$ with rod vector $\partial_\psi$; note that the $e^{\tilde{W}} \hat{C}$ and $(1-y^2) e^{2\tilde{W}}$ are smooth and non-zero on this rod and hence $g_{\phi\phi}>0$ there, and (2) a semi-infinite rod $y =1$ with $x > \lambda>1$ with rod vector $\partial_\phi$ provided we fix
 \begin{figure}[!ht]
\centering
\includegraphics[scale=0.25]{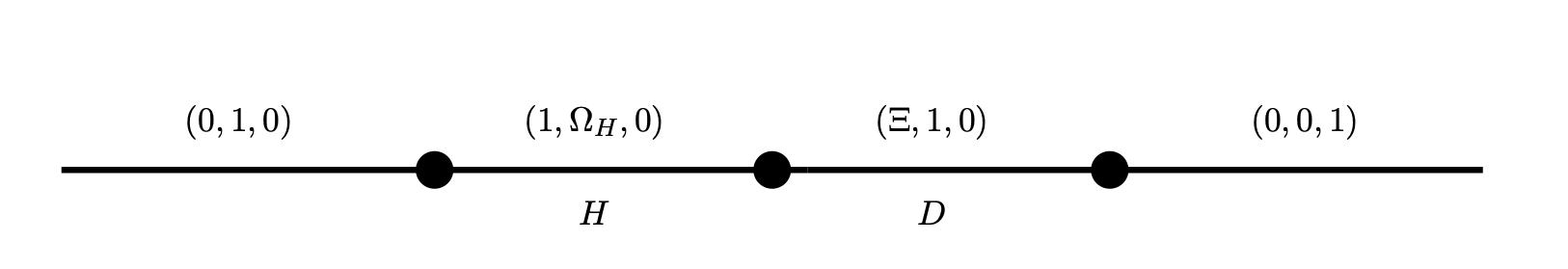}
\caption{Rod data set for the local solution. The black ring occurs when $\Xi=0$, so the rods adjacent to the horizon rod $H$ have the same rod vector. The two-dimensional disc $y=-1, x \in (1,\lambda)$ is indicated.}
\end{figure}
\begin{equation}
    {c}_1 = \frac{2\alpha \sqrt{\sigma}}{1 + \alpha \beta}. 
\end{equation} This ensures that the undistorted solution is asymptotically flat.  There is a finite horizon rod $x =1, -1 < y < 1$ upon which the Killing vector field
\begin{equation} \label{nullgen}
    \xi = \frac{\partial}{\partial t} + \Omega_H \frac{\partial}{\partial \phi}, \qquad \Omega_H = \frac{(1 + \alpha \beta)(\alpha(1 + \lambda) - 2\beta)}{2 \sqrt{\sigma} (2 + \alpha^2(1 + \lambda))}. 
\end{equation} is null. Finally, on the finite rod $y = -1, 1 < x < \lambda$ the following spatial Killing field that degenerates is
\begin{equation}
 \zeta =    \Xi \frac{\partial}{\partial t} + \frac{\partial}{\partial \phi}, \qquad \Xi = \frac{2\sqrt{\sigma} (2 \beta + \alpha(\lambda -1 + \beta(2\beta + (1+\lambda) \alpha))}{(1 + \alpha \beta)(\lambda - 1 + (1+ \lambda)\alpha \beta)}.
\end{equation} To ensure that the horizon has topology $S^1 \times S^2$, we require that the same Killing vector field vanishes on the two rods adjacent to the horizon rod, namely $\partial_\phi$. This can be achieved provided $\Xi =0$, which requires the parameter $\beta$ to satisfy a quadratic equation. The choice of $\beta$, which can be distorted to produce a regular solution free of conical singularities is
\begin{equation}\label{undistortedbeta}
    \beta = - \frac{1}{4\alpha} \left( 2 + (1 + \lambda)\alpha^2 - \left[8\alpha^2( 1 - \lambda) + (2 + (1 + \lambda)\alpha^2)^2\right]^{1/2}\right).
\end{equation}

\subsubsection{Distorted solution}
For the distorted case, note that as observed above, the location of the rods which is characterized by the zero sets of $\hat\rho$ on the $z-$ axis is the same as that of the undistorted case. However, the lengths of the rods and rod vectors, which control which linear combination of Killing fields vanish, do indeed change. This can be seen directly from the local metric \eqref{metricxy} by the fact that the distortion functions $e^{\widehat{U}}, e^{\widehat{W}}$ are bounded but do not vanish. In particular, to maintain the rod structure on the semi-infinite rods, we again must choose the constant $\hat{c}_1$ appropriately so that $\partial_\phi$ degenerates on $y=1, x > \lambda$. To preserve the horizon topology to be a ring, we require that the Killing vector $\zeta = \partial_\phi$; this imposes $\Xi =0$. Finally, by demanding that a timelike Killing field $\xi$ becomes null on the horizon rod, we can determine the constant $\Omega_H$. 

For an arbitrary distortion (i.e. arbitrary $(a_n,b_n)$), the expressions are somewhat cumbersome and not particularly useful. We will focus on fixing attention to particular distortions (e.g. dipole distortions with only $a_1, b_1$ non-zero). 

Firstly, on the semi-infinite rod $y=1$, $x > \lambda > 1$ the requirement that $\partial_\phi$ degenerates for any fixedx non-zero odd multiple moments, $n=1,3,..$
\begin{equation}
    \hat{c}_{1}= \frac{2\alpha\sqrt{\sigma}\exp{\left(-2a_{n}-b_{n}\right)}}{\alpha \beta \exp{\left(-4a_{n}-2b_{n}\right)}+ 1 }
\end{equation} whereas for a fixed non-zero even multiple moment $m=2,4,...$ we get
\begin{equation}
  \hat{c}_1 = \frac{2\alpha \sqrt{\sigma}\exp{(-2a_{m}-b_{m})}}{\alpha \beta+1}.    
\end{equation} On the finite horizon rod $x =1, -1 < y < 1$ for the distorted rotating black ring, the angular velocity for odd multiple moments $n=1,3,...$ (that is, all even multiple moments, $a_j, b_j$, vanish except possibly $a_n, b_n$): 
\begin{equation}
 \Omega_H = \frac{e^{(-2a_n-b_n)}(e^{(4a_n+2b_n)} + \alpha \beta)(\alpha(1 + \lambda) - 2\beta e^{(-4a_n-2b_n)})}{2 \sqrt{\sigma} (2 + \alpha^2(1 + \lambda))},
\end{equation} whereas for even multiple moments $m=2,4,...$
\begin{equation}
    \Omega_H = \frac{e^{(2a_m+b_m)}(1 + \alpha \beta)(\alpha(1 + \lambda) - 2\beta)}{2 \sqrt{\sigma} (2 + \alpha^2(1 + \lambda))}. 
\end{equation} 
 On the finite rod $y = -1, 1 < x < \lambda$ for the distorted case, $n=1$, requiring $\zeta = \partial_\phi$ to degenerate fixes, for odd multiple moments $n=1,3,...$
\begin{equation}
    \Xi =\frac{2 \sqrt{\sigma} e^{2 a_n + b_n}\left((2 + \alpha^2 (1 + \lambda)) \beta e^{4 a_n + 2 b_n}
+ \alpha \left( 2\beta^2 + \lambda -1 \right)  \right)}
{((1 + \lambda) \alpha^2 \beta^2 + \lambda - 1) e^{4 a_n+ 2 b_n}
+ \alpha \beta \left( (1 + \lambda) e^{8 a_n + 4 b_n} + \lambda - 1 \right)},
\end{equation} whereas for even multiple moment $m=2,4,...$
\begin{equation}
    \Xi = \frac{2\sqrt{\sigma} e^{-2a_m-b_m}(2 \beta + \alpha(\lambda -1 + \beta(2\beta + (1+\lambda) \alpha))}{(1 + \alpha \beta)(\lambda - 1 + (1+ \lambda)\alpha \beta)}.
\end{equation} The conditions on $\Omega_H$, $\Xi$, and $\hat{c}_1$ reduce to those given in Section \ref{undistorted} when the distortions are removed. The vanishing of $\Xi$ fixes two roots $\beta = \beta_\pm$. It will turn out that only the odd multiple moments will be relevant for removing conical singularities. 
For any odd multiple moments, $n^{\prime}=1,3,..$
\begin{eqnarray}
    \beta_{-}=  \frac{1}{4 \alpha} \bigg(\sqrt{
e^{\sum_{n^{\prime}=1}^{\infty}(8a_{n^{\prime}}+4b_{n^{\prime}})}\left[\alpha^{2}(\lambda+1)+2\right]^{2}
		- 8 \alpha^2(\lambda-1)
}-e^{\sum_{n^{\prime}=1}^{\infty}(4a_{n^{\prime}}+2b_{n^{\prime}})}\left(\alpha^2 (\lambda +1)+2 \right)
\bigg)
\end{eqnarray}
\begin{eqnarray}
	\beta_{+}= -\frac{1}{4 \alpha} \bigg( \sqrt{
		e^{\sum_{n^{\prime}=1}^{\infty}(8 a_{n^{\prime}}+4 b_{n^{\prime}})}\left[\alpha^{2}(\lambda+1)+2\right]^{2}
		- 8 \alpha^2(\lambda-1)
	} + e^{\sum_{n^{\prime}=1}^{\infty}(4 a_{n^{\prime}}+2 b_{n^{\prime}})} \left(\alpha^2 (\lambda +1)+2 \right)
	\bigg)
\end{eqnarray} In particular, for the dipole case, where $a_{1},b_{1} \neq 0$ and all of the other multiple moments are equal to zero, we obtain
\begin{eqnarray}
\beta_{-}= \frac{1}{4 \alpha} \bigg( \sqrt{
	e^{8 a_1+4 b_1}\left[\alpha^{2}(\lambda+1)+2\right]^{2}
	- 8 \alpha^2(\lambda-1)
} - e^{4 a_1+2 b_1} \left(\alpha^2 (\lambda +1)+2 \right)
\bigg) , 
\end{eqnarray}
\begin{eqnarray}
	\beta_{+}= -\frac{1}{4 \alpha} \bigg( \sqrt{
		e^{8 a_1+4 b_1}\left[\alpha^{2}(\lambda+1)+2\right]^{2}
		- 8 \alpha^2(\lambda-1)
	} + e^{4 a_1+2 b_1} \left(\alpha^2 (\lambda +1)+2 \right)
	\bigg).
\end{eqnarray} Note that $\beta_-$ reduces to \eqref{undistortedbeta} when the distortions are set to zero. One can show that, in the undistorted case, this choice guarantees that $g_{\phi\phi} \geq 0$ everywhere in the domain of outer communication \cite{iguchi2006}. By continuity, this should extend to the distorted case, at least for sufficiently small distortions. Therefore, we will assume $\beta = \beta _-$ in the remainder of this work. 
\subsection{Regularity}
Having established the rod structure, we now turn to the issue of smoothness of the metric. The metric functions are smooth away from the set $\rho =0$. We wish to remove the conical singularities that arise where a rotational Killing field fails to degenerate smoothly. Such singularity necessarily arises in the undistorted case and is interpreted as meaning that an external force is required to prevent collapse of the black ring (that is, angular momentum along the $S^2$ of the ring is insufficient to prevent collapse). Removing conical singularities will impose restrictions on the (up to now undetermined) periodicities of the angular coordinates $(\psi,\phi)$. Recall that if $\eta = v^i \partial_i$ with $\partial_i = (\partial_\psi,\partial_\phi)$ degenerates on an interval $\rho = 0, z \in I$, the removal of conical singularities requires that the period $\Delta \eta$ of the orbits of $\eta$ is fixed as
\begin{equation}
    \Delta \eta = 2\pi \lim_{\rho \to 0} \sqrt{\frac{e^{2\nu} \rho^2}{g_{ij}v^i v^j}}
\end{equation} where $e^{2\nu} = g_{\rho\rho} = g_{zz}$ can be read off from \eqref{metricrz1}. On the semi-infinite rod $y =-1$, $x > \lambda$, $\partial_\psi$ degenerates (that is, $v^i = (1,0)$), fixing $\Delta \psi = 2\pi$ can be used to determine the integration constant $\hat{c}_2$.
For any odd multiple moments, $n=1,3,..$
\begin{equation}
    \hat{c}_{2} = \frac{\sqrt{2}\exp{(\lambda^{n} (b_{n} - a_{n}) +4a_{n} +2b_{n})}}{2\left((\alpha\beta) \exp{(4a_{n}+2b_{n})}+1\right)^{2}},
\end{equation} whereas for even multiple moments, $m=2,4,..$
\begin{equation}
    \hat{c}_{2} = \frac{\sqrt{2}\exp{(\lambda^{n} (b_{n} - a_{n}))}}{2\left(\alpha\beta+1\right)^{2}}.
\end{equation} We now turn to the determination of the period of $\phi$. For the range $1 < x < \lambda $, in the dipole case, where $a_{1},b_{1} \neq 0$ and the other multiple moments vanishing, we obtain
\begin{equation}
		\Delta \phi_{1}=2\pi\left(\dfrac{\lambda-1+(\lambda+1)\alpha \beta e^{4a_{1}+2b_{1}}}{\sqrt{\lambda^2-1}(e^{-(4a_{1}+2b_{1})}+\alpha\beta)}\right)e^{\lambda(b_{1}-a_{1})-4a_{1}-2b_{1}}
\end{equation}
If $a_{3},b_{3} \neq 0$ and for $n\neq 3$, $a_{n}=b_{n} = 0$, we have
\begin{equation}
\Delta\phi_{1} = 2\pi \sqrt{
\frac{
e^{\left(-{a_3}+ {b_3}\right)\lambda^3 -4a_3 -2b_3}
\left((2\lambda^2 - 2)\,\beta\alpha\, e^{4a_3+2b_3} + \beta^2 \alpha^2 (1+\lambda)^2 e^{8a_3+4b_3} + (\lambda -1)^2\right)
}
{
\left(e^{\left(a_{3}- b_{3}\right)\lambda^3 +4a_3 +2b_3}\alpha^2 \beta^2
+ 2e^{\left(a_{3}- b_{3}\right)\lambda^3}\alpha \beta
+ e^{\left(a_{3}-b_{3}\right)\lambda^3 -4a_3 -2b_3}\right)(\lambda^2 -1)
}
}.
\end{equation}
For the semi-infinite rod $ y=1,x>1 $, in the dipole case where $a_{1},b_{1} \neq 0$ and for $n\neq 1$, $a_{n}=0$, we obtain
\begin{equation}
	\Delta \phi_{2}=2\pi\left(\dfrac{\alpha \beta e^{-(4a_{1}+2b_{1})}+1}{(e^{-(4a_{1}+2b_{1})}+\alpha\beta)}\right)e^{\lambda(b_{1}-a_{1})}
\end{equation} and if $a_{3},b_{3} \neq 0$ and for $n\neq 3$, $a_{n}=b_{n} =0$, we have
\begin{equation}
    \Delta\phi_{2} := 2\pi\sqrt{
\frac{
\left(\alpha^2\beta^2 e^{-8a_{3}-4b_{3}} + 2\alpha\beta e^{-4a_{3}-2b_{3}} + 1\right)
e^{\left(-a_{3}+b_{3}\right)\lambda^3 +4a_{3}+2b_{3}}
}{
e^{\left(a_{3}-b_{3}\right)\lambda^3 +4a_{3}+2b_{3}}\alpha^2\beta^2
+ 2 e^{\left(a_{3}-b_{3}\right)\lambda^3}\alpha\beta
+ e^{\left(a_{3}-b_{3}\right)\lambda^3 -4a_{3}-2b_{3}}
}
}.
\end{equation}
Global regularity requires that these periodicities are consistent,  that is, imposing $	\Delta \phi_{1}=\Delta \phi_{2}  $. The regularity condition for the dipole case requires
\begin{equation}\label{rega1b1}
	a_{1} = -\frac{b_{1}}{2} -\frac{1}{4} \ln \sqrt{\dfrac{(\lambda + 1)}{(\lambda - 1)}}.
\end{equation}
If $a_{3},b_{3} \neq 0$ and for $n\neq 3$, $a_{n}=0$, the regularity condition is
\begin{equation}
a_{3}=-\frac{b_3}{2} -\frac{1}{4} \ln \sqrt{\dfrac{(\lambda + 1)}{(\lambda - 1)}}.
\end{equation} In general, for an arbitrary superposition of odd multiple moments, $n^{\prime}=1,3,..n$, the regularity condition is
\begin{equation}
\sum_{n^{\prime}=1}^{n}\left(a_{n^{\prime}}+\frac{b_{n^{\prime}}}{2}\right)+\frac{1}{4} \ln \sqrt{\dfrac{(\lambda + 1)}{(\lambda - 1)}}=0.
\end{equation}
We note that even-mode distortions (e.g. quadrupole) do not contribute towards removing conical singularities. This is because the even modes leave the metric invariant under a reflection symmetry $y \to -y$ which is incompatible with imposing $\Delta \phi_1 = \Delta \phi_2$ (recall that $\partial_\phi$ degenerates at $y = \pm 1$). 

\section{Properties of the solution}
\begin{figure}
 \centering
 \includegraphics[width=7.5cm]{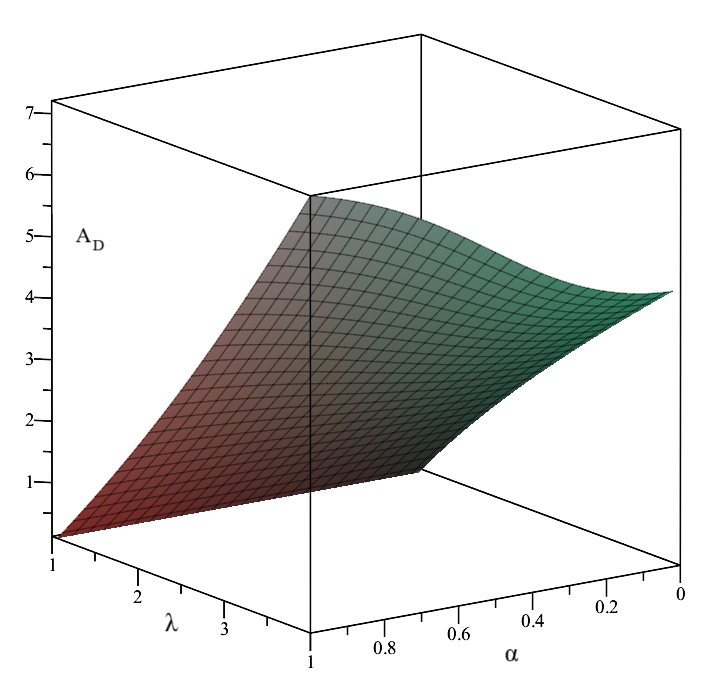}~~~
	\caption{Area of the disk for an undistorted rotating black ring normalized with $\sigma =1$.}
	\label{A_D-undistortedRBR}
\end{figure}

\begin{figure}
 \centering
 \includegraphics[width=7.5cm]{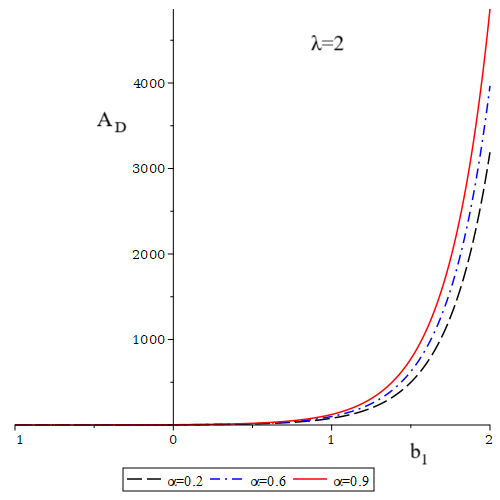}~~~
 \includegraphics[width=7.5cm]{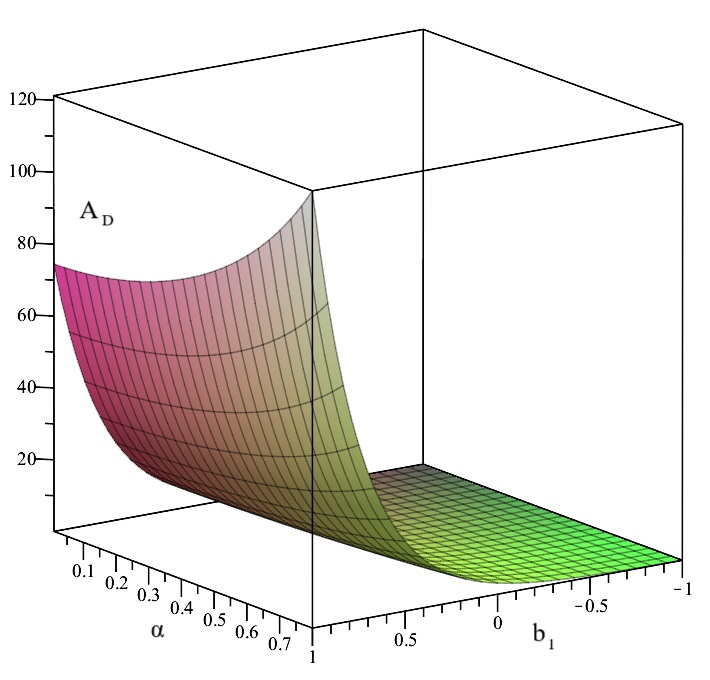}~~~
	\caption{ The illustrations of the area of the disk for a distorted rotating black ring for different values of $\alpha$, $ b_1$, $\sigma=1$, and $\lambda=2$.}
	\label{A_D-distortedRBR}
\end{figure}

\begin{figure}
 \centering
 \includegraphics[width=7.5cm]{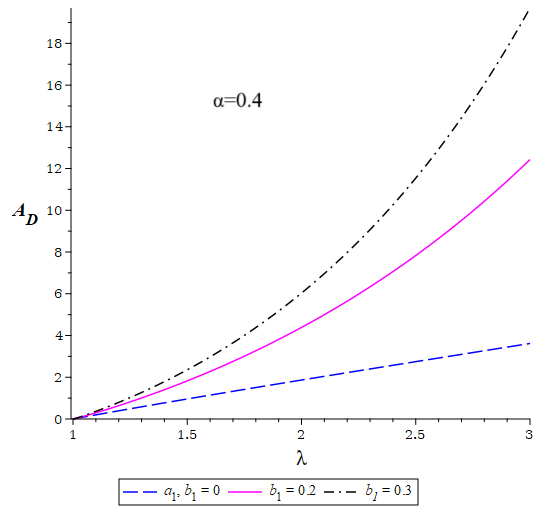}~~~
	\caption{The behaviour of the disk area of a rotating black ring for both undistorted and distorted cases. The blue line represents the area of an undistorted disk.}
	\label{A_D-comparison}
\end{figure}

The distortions preserve the event horizon of the original solution, although its geometry is altered and made smooth without conical singularities. In the following, we will consider the simplest set of distortions that are sufficient to achieve regularity, namely, with all $(a_n,b_n)$ vanishing except $a_1, b_1 \neq 0$. Consider first spatial sections of the event horizon, defined by $t=\const$ and $x=1$. On the horizon
\begin{equation}
    \hat{a}= 
\frac{
  \sqrt{2} \alpha (\lambda+1) e^{-(2a_{1} + b_{1})(y-1)} 
}{
   2 \sqrt{y+\lambda}
}
\end{equation}
\begin{equation}
  \hat{b}= \frac{
    \sqrt{2} \beta e^{-(2a_{1} + b_{1}) (y+1)
}}{
    \sqrt{y+\lambda}
}
\end{equation}
The metric on the horizon is
\begin{equation}
\begin{aligned}
g_{\phi \phi}&=\frac{
  4 \sigma (\alpha^2 \lambda + \alpha^2 + 2)^2 e^{2(a_{1}+b_{1})y+4(2a_{1}+b_{1})} \, (1-y^{2})
}{W(y)}  \\
W(y) & =
    \left( e^{4a_1 + 2b_1} + \alpha \beta \right)^2
    \left( 8(y+\lambda) e^{(4a_1 +2b_1)y}
        +  \alpha^2 (1 + \lambda)^2 (1 + y)^2 e^{4a_1 + 2b_1} \right.\\ 
        &\left. - 4\beta(y-1) \left(
          \alpha(1 + \lambda)(1 + y)+(1-y)e^{(-4a_1-2b_1)}\beta
        \right)
    \right)
\end{aligned}
\end{equation}
\begin{equation}
    g_{\psi\psi}=2\sigma(\lambda+y)e^{(-2b_{1}y)}
\end{equation}
\begin{eqnarray}
   g_{yy}&=& -\frac{\sigma e^{(-2a_1 + 2b_1)\lambda - 2a_1 y}\left[8(\lambda +y)e^{2(2a_1 + b_1)(1 + y)}
        +\alpha^2 (1 + \lambda)^2 (1 + y)^2 e^{8a_1 + 4b_1}\right] }{(1 + \lambda) \left( \alpha \beta e^{4a_1 + 2b_1} + 1 \right)^2(\lambda + y)(y^2 - 1)}\\ \nonumber
  &&   +\frac{ \sigma \beta e^{(-2a_1 + 2b_1)\lambda - 2a_1 y}\left[\alpha (1 +\lambda)(1-y^2) e^{4a_1 + 2b_1} + \beta(y-1)^2\right]}{(1 + \lambda) \left( \alpha \beta e^{4a_1 + 2b_1} + 1 \right)^2(\lambda + y)(y^2 - 1)}
\end{eqnarray}
If we consider the angular coordinate $0\leq\theta\leq \pi$ such that $y=\cos\theta$,
\begin{eqnarray}
    g_{\phi\phi}=\dfrac{\sigma (\sin\theta)^2\left( \alpha^2 \lambda + \alpha^2 + 2 \right)^2 e^{\left(2 a_1 + 2 b_1 \right) \cos\theta + 12 a_1 + 6 b_1}  
}{\left(e^{(4a_{1}+2b_{1})}+\alpha\beta\right)^2\left(2(\lambda+\cos\theta)e^{4(\cos(\theta/2))^{2}(2a_{1}+b_{1})}+\left[(\cos(\theta/2))^{2}\left(\alpha(1+\lambda)e^{4a_{1}+2b_{1}}-2\beta\right)+2\beta\right]^2\right)}
\end{eqnarray}
\begin{align}
g_{\theta\theta} &=\dfrac{\sigma e^{-2\lambda(a_{1}-b_{1})-2a_{1}\cos\theta}\left(2(\lambda+\cos\theta)e^{4(\cos(\theta/2))^{2}}+\left[(\cos(\theta/2))^{2}\left(\alpha(1+\lambda)e^{4a_{1}+2b_{1}}-2\beta\right)+2\beta\right]^2\right)}{(1+\lambda)\left(e^{(4a_{1}+2b_{1})}\alpha\beta+1\right)^2 (\sin\theta)^2\left(\lambda+\cos\theta\right)}
\end{align}

\begin{equation}
 g_{\psi\psi}=2\sigma(\lambda+\cos\theta)e^{(-2b_{1}\cos\theta)}
\end{equation}
The horizon area is defined as 
\begin{equation}
    \mathcal{A}_{_H} =\int_H d\text{vol}(g_H) = 4\pi^2 \int_{-1}^1 \sqrt{\det g_H} \; dy = 4\pi^2 \int_{0}^\pi \sqrt{\det g_H} \; d\theta
\end{equation}
After calculating this integral, we obtain
\begin{equation}
    \mathcal{A}_{_H}= \frac{8\pi^2 \sqrt{2} \left(2 + (1 + \lambda)\alpha^2\right) \sigma^{3/2} 
e^{(b_1 - a_1)\lambda + 6a_1 + 3b_1}}
{\sqrt{1 + \lambda} \left( \left( \alpha^2 \beta^2 + 1 \right) e^{4a_1 + 2b_1} + \left( e^{8a_1 + 4b_1} + 1 \right)\beta \alpha \right)}
\end{equation} where $b_1$ is fixed by the condition \eqref{rega1b1} which removes concial singularities. 
There is a non-collapsed two-dimensional disc region that ends on the horizon, which is represented in the $(x,y)$ coordinates as the region $y = -1, 1 < x < \lambda$. On this disc, the Killing field $\partial_\phi$ degenerates (since $\Xi =0$) and spatial sections have the geometry
\begin{equation}
    g_D = g_{xx} dx^2 + g_{\psi\psi} d\psi^2
\end{equation} and it has area
\begin{equation}
    A_D = \int_D d\text{vol}(g_D) = 2\pi \int_1^\lambda \sqrt{\det g_D} \, dx
\end{equation}
Due to the complexity, we have been unable to explicitly carry out this integral. Instead, we computed it numerically for various parameter values. Figure \ref{A_D-undistortedRBR} shows the disk area of a five-dimensional rotating black ring as a function of the parameters $\alpha$ and $\lambda$ in the absence of any external distortion. The disk area 
 increases with both these parameters, but the dependence on $\lambda$ is more dominant. The behaviour of the disk area of a five-dimensional rotating black ring in the presence of external gravitational fields is shown in Figure \ref{A_D-distortedRBR}.
As $b_{1}$ increases, the disk area grows rapidly, showing that the external field stretches the disk.
Increasing $\alpha$ tends to increase the disk area.
This indicates that rotation enhances the inflationary effect of the external distortion.
Figure \ref{A_D-comparison} provides a comparative analysis of how the disk area of a five-dimensional rotating black ring evolves with the parameter $\lambda$ and for different values of the external field parameter $b_1$. The blue curve represents the disk area of a rotating black ring without the presence of external gravitational fields. We see that even moderate external fields begin to stretch the disk geometry, and increasing $\lambda$ leads to an increase in the disk area. Furthermore, the difference between the curves widens rapidly as $\lambda$ increases, indicating external fields have a stronger impact on more extended rings. In the limit $\lambda \to 1$, the black ring solution reduces to the five-dimensional Myers–Perry black hole with a single rotation. Correspondingly, the disk area 
smoothly tends to zero in this limit. As observed in figure \ref{A_D-comparison}, all curves intersect at the origin point, and this reflects the fact that in the Myers–Perry geometry, there is no disk area to be distorted and the external dipole fields become irrelevant in this limit.


For an asymptotically flat solution, the ADM mass and angular momentum are well defined. For the undistorted solution, by reading off the asymptotic form of $g_{tt}$, given by
\begin{equation}
    g_{tt} = -\left(1 - \frac{4\sigma(1 + \alpha^2(1 - \beta^2))}{(1+ \alpha \beta)^2 r^2} \right) + O(r^{-4})
\end{equation} where $r$ is a standard radial coordinate in the asymptotic region $\mathbb{R}^4 \setminus $Ball related to the canonical coordinates $(\rho,z)$ by $r = \sqrt{2} (\rho^2 + z^2)^{1/4}$, we can read off the mass from the expansion $g_{tt} = -1 + 8M /(3\pi r^2) + O(r^{-4})$:
\begin{equation}\label{ADMmass}
    M = \frac{3 \pi}{2} \cdot \frac{\sigma(1 +\alpha^2(1 - \beta^2))}{(1+ \alpha \beta)^2}.
\end{equation} The non-vanishing angular momentum in the $\phi-$ direction can similarly be read off from expansion $g_{t\phi} = 4J_\phi \sin^2\theta/(\pi r^2) (1 + O(r^{-2})$, giving
\begin{equation}
    J_\phi = \frac{\pi \sigma^{3/2}}{(1+ \alpha \beta)^3} \left[ \alpha ( 2\alpha^2 - \lambda + 3) - 2 \alpha^2 \beta^3 - \beta(2(2\alpha \beta + 1)(1 + \alpha^2) - (1-\lambda)\alpha^2(2 + \alpha \beta))\right]
\end{equation} 
The surface gravity $\kappa_H$ is defined by $\xi^b \nabla_b \xi^a = -2\kappa_H \xi^a$ where the evaluation is taken on the horizon. A standard calculation shows that it can be obtained from 
\begin{equation}
    \kappa_H = \lim_{x\to 1} \sqrt{-\frac{g^{ab} \partial_a |\xi|^2 \partial_b |\xi|^2}{4 |\xi|^2}}
\end{equation} where $|\xi|^2 = g_{ab}\xi^a \xi^b$ is the inner product of the null generator of the horizon given by \eqref{nullgen}. For the undistorted case, we find
\begin{equation}
    \kappa_H = \frac{(1 + \alpha\beta)^2 \sqrt{1 + \lambda}}{\sqrt{2 \sigma} (2 + (1 + \lambda)\alpha^2)}
\end{equation} In the undistorted asymptotically flat setting, the charges satisfy
\begin{equation}
    M = \frac{3}{16\pi} \kappa_H A_H + \frac{3}{2} \Omega_H J_\phi.
\end{equation} where we have used \eqref{undistortedbeta}. This Smarr relation (which appears in \cite{Figueras} using a completely different parametrization), is derived from integrating a divergence-type identity, and hence is not sensitive to the conical singularity inherent to the undistorted solution. However, the first law, which is a differential identity, will not hold. A similar situation holds for the family of static black ring solutions \cite{EmparanReall}.

For the distorted case, the solution is, of course, not asymptotically flat, and hence the ADM mass and angular momentum are not defined. Alternatively, one may use a Komar-type prescription to define conserved changes associated with time translations and rotations (these agree with the ADM definitions in the stationary and axisymmetric case). However, physical quantities defined in this way are only defined by an overall normalization factor (i.e. the generator can be scaled arbitrarily). In the asymptotically flat case, this factor is fixed by the requirement that the generators agree with those of Minkowski spacetime. However, in the distorted case, this is not possible, resulting in an ambiguity in defining the mass, angular momentum, and surface gravity. 

In more detail, we can define quasi-local Komar charges evaluated on the horizon: 
\begin{equation}
    M_H = -\frac{3}{32\pi} \int_H \star d k, \qquad J_H = \frac{1}{16\pi} \int_H \star d \zeta
\end{equation} where $k = g(\partial_t,~)$ and $\zeta = g(\partial_\phi,~)$ are the one-forms dual to the Killing fields generating time translations and rotations (as noted above, these can be arbitrarily scaled by a non-zero constant). It is a simple application of Stokes' theorem and Ricci flatness to show that the above integrals are equal to the corresponding quantities defined at spatial infinity in the asymptotically flat setting, and hence these are reasonable candidates to define the mass and angular momentum in the distorted case. Unfortunately, we haven't been able to compute these horizon integrals due to the complexity of the metric on the horizon. It would be interesting to pursue this further in order to verify whether the first law does indeed hold. 

\begin{figure}
 \centering
 \includegraphics[width=7.5cm]{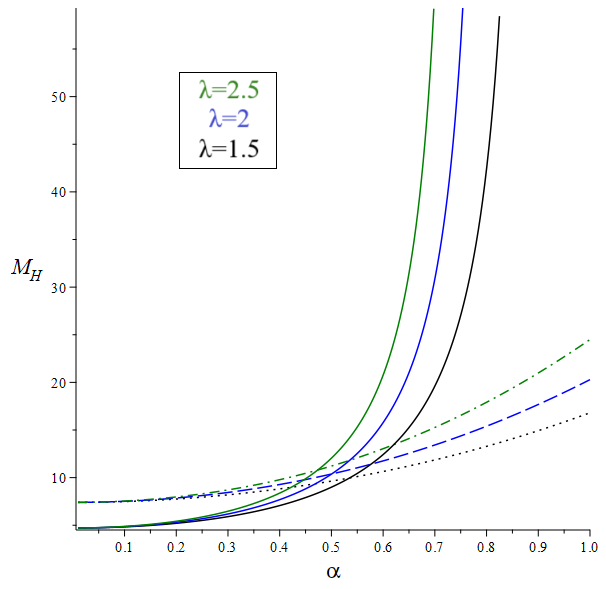}~~~
	\caption{The behaviour of Komar mass on the horizon of a rotating black ring for distorted (solid curves) and undistorted (dashed/dotted curves) cases for different values of $\lambda$.}
	\label{MH-RBR}
\end{figure}

In Figure \ref{MH-RBR}, the variation of the Komar mass with the parameter $\alpha$ is illustrated by computing it numerically for various parameter values.
This figure illustrates the dependence of the Komar $M_H$ on $\alpha$ for a rotating black ring, comparing distorted (solid curves) and undistorted (dashed/dotted curves) configurations for different values of $\lambda$. It is worth mentioning that this distorted black ring is free of conical singularities, and the regularity condition of \ref{rega1b1} is applied to the distorted solution.
In general, it can be seen that Komar mass increases with the parameter $\alpha$, starting from a finite minimum at small $\alpha$. Furthermore, the rate of increase of Komar mass accelerates as $\alpha$  grows. This behaviour of the Komar mass is in agreement with the ADM mass expression in \ref{ADMmass}, both increase with the parameter $\alpha$. 
For small values of $\alpha$, all curves nearly coincide, indicating that the effect of $\lambda$ on the Komar mass is negligible. As $\alpha$ increases, the curves begin to diverge, with larger $\lambda$ producing a steeper growth in $M_H$.
Interestingly, the solid curves (distorted cases) rise more steeply than their undistorted counterparts, with the effect becoming more pronounced for larger $\lambda$.
It would be interesting to pursue this further in order to verify whether the first law does indeed hold. 
\section{Conclusion}
In this paper, we have constructed a new solution representing a distorted rotating black ring. Through the careful synthesis of solitonic solution-generating techniques, the generalized Weyl formalism, and the theory of distorted black holes, we can demonstrate that it is possible to obtain smooth, stationary, asymptotically non-flat vacuum rotating black hole solutions with horizon topology $S^1\times S^2$ that are free from conical singularities. The distorted black ring rotates around the $S^2$ direction. A distorted black hole represents a `local' solution sufficiently far away from the external sources and interior to the sources.  This work examines the interplay between geometry, external fields, and horizon structure in a particular higher-dimensional setting. While the generalized Weyl formalism has proven useful for generating static, axisymmetric solutions by embedding external multipole distortions, it falls short in stationary non-static settings. To overcome this, we turned to the B\"acklund transformation, which allowed us to generate rotating solutions from a suitably chosen static, distorted seed. In this case, we employed a distorted five-dimensional Minkowski spacetime modified by axisymmetric multipole moments as the seed geometry. 
One of the central challenges in constructing rotating black rings lies in achieving regularity, i.e. a solution that is free of conical singularities. Only the presence of the odd multiple moments allows for the removal of the conical singularity. By carefully incorporating specific external multipole distortions, we were able to remove the conical singularity. The black ring horizon remains regular and admits a well-defined null generator with well-defined surface gravity and angular velocity. It would be interesting to apply our approach to other rotating black hole solutions with non-spherical topology that also suffer from geometric pathologies. For example, Chen and Teo have constructed a rotating, asymptotically flat vacuum black hole solution with lens-space horizon topology $L(2,1) = S^3 / \mathbb{Z}_2$. Members of this family of solutions necessarily suffer from a conical singularity along one of the axes of rotation \cite{ChenTeo} and it has been shown that this pathology must hold for any vacuum, asymptotically flat black lens~\cite{Lucietti:2020phh}. It would be of interest to investigate whether one can produce regular local black lens solutions of this type.

\paragraph{Acknowledgements}  HKK acknowledges the support of the NSERC Discovery Grant 2025-06027. I.B. acknowledges the support of NSERC Discovery Grant 2018-04873. M.T. was supported by both of
these grants. S.A. acknowledges that this research was supported in part by the grant NSF PHY-1748958 to the Kavli Institute for Theoretical Physics (KITP).

%

\end{document}